\documentclass{nature_modified_OP}
\usepackage[scaled=1]{helvet}          

\usepackage{lineno}

\usepackage{graphicx}      
\usepackage{booktabs}      
\usepackage[section]{placeins} 
\graphicspath{{figures/}}

\usepackage{xcolor}
\usepackage{newfloat} 
\usepackage{amsmath}
\usepackage{amssymb}
\usepackage{soul}
\usepackage{cite}       
\usepackage{microtype}  
\usepackage{caption}
\DeclareCaptionLabelSeparator{bar}{ \textbar\ }
\DeclareCaptionFont{ninept}{\linespread{1}\fontsize{9}{11}\selectfont}
\DeclareFloatingEnvironment[name={Extended Data Figure}]{suppfigure}

\makeatletter
\renewenvironment{figure}{\@float{figure}[htbp]}{\end@float}
\renewenvironment{table}{\@float{table}[htbp]\sffamily}{\end@float}
\makeatother

\title{Rapid Detection of Neurological Anomaly using Deep Representation Learning of Normative Molecular MRI}

\author{Sahar Ifrah$^1$, Adi Eliyahu$^2$, Michal Rivlin$^1$, Ruth Ben Chaim$^3$, Hagar Shmuely$^1$, Or Zohar$^4$,   Dinorah Friedmann-Morvinski$^{4,3}$, Simon Weinmüller$^{5}$, Moritz Zaiss$^{5,6}$, Shelly Gur Zeldich$^{7,8}$, Dafna Ben Bashat$^{3,7,8}$, Shlomit Yust-Katz$^{8,9}$, Rotem Sivan-Hoffman$^{10}$, Or Perlman$^{1,3*}$}

\begin{document}

\maketitle

\begin{affiliations}
 \item School of Biomedical Engineering, Tel Aviv University, Tel Aviv, Israel
 \item Faculty of Health Sciences, Ben-Gurion University of the Negev, Beer-Sheva, Israel
 \item Sagol School of Neuroscience, Tel Aviv University, Tel Aviv, Israel
 \item School of Biochemistry, Neurobiology, and Biophysics, Tel-Aviv University, Tel-Aviv, Israel
 \item Institute of Neuroradiology, University Hospital Erlangen, Friedrich-Alexander Universität Erlangen-Nürnberg, Erlangen, Germany
\item Department of Artificial Intelligence in Biomedical Engineering, Friedrich-Alexander Universität Erlangen-Nürnberg, Erlangen, Germany
\item Sagol Brain Institute, Tel Aviv Sourasky University Medical Center, Tel Aviv, Israel
\item Gray Faculty of Medical and Health Sciences, Tel Aviv University, Tel Aviv, Israel
\item Neuro-Oncology Unit, Rabin Medical Center, Beilinson Hospital, Petach Tikva, Israel
\item Radiology Department, Meir Medical Center, Kfar Saba, Israel 
\end{affiliations}

\noindent{$^*$Correspondence to: Or Perlman, School of Biomedical Engineering and Sagol School of Neuroscience, Tel Aviv University, Tel Aviv, Israel. Email: orperlman@tauex.tau.ac.il}\\


\begin{abstract}
\textbf{Abstract}\\
The sharp increase in magnetic resonance imaging (MRI) examinations over the last decade has opened new opportunities for improved medical diagnostics, but has also substantially increased the radiologist workload. Here we show that multiple sclerosis, Parkinson's disease, brain cancer, and aging effects can be automatically detected in mice (n=82, 22\% test set) using an unsupervised anomaly detection (UAD) framework that learns the normative distribution of molecular MRI. 
The framework fuses forward-diffusion noising with a saturation transfer excitation-informed vision transformer to achieve an area under the receiver operating characteristic curve (AUROC) of 0.961 [95\% confidence interval (CI): 0.87-1.00]  and an area under the precision-recall curve (AUPRC) of 0.982 [95\% CI: 0.94-1.00] when fed rapidly acquired semisolid magnetization transfer (MT) image data. Translating the method to clinical scanners at four sites (n=36 subjects, 53\% test set), provided an AUROC of 0.964 [95\% CI: 0.86-1.00] and an AUPRC of 0.981 [95\% CI: 0.93-1.00] for human subject-specific detection of brain cancer and Alzheimer's disease. The framework requires less than 43 seconds for MR data acquisition and less than 6 seconds for inference and can be extended to additional anomalies, to enable automatic flagging of urgent cases and prioritization in clinical workflows.

\end{abstract}
 
\section*{Introduction}

\noindent Magnetic resonance imaging (MRI) is playing an increasingly important role in clinical diagnostics, with some countries seeing more than a two-fold rise in the numbers of examinations over the past decade\cite{OECD2025}. This growth has increased the interpretive burden on radiologists, leaving less time for each image evaluation and raising the risk of missing pathological findings\cite{alexander2022mandating}.
The multidisciplinary MRI method development community (physicists, physicians, chemists, and engineers) has brought several new contrast modalities to approval by the U.S. Food and Drug
Administration (FDA), incorporation into routine neurological diagnostic criteria\cite{montalban2025diagnosis}, and consensus recommendations\cite{zhou2022review}. This opens new opportunities for improved diagnostic accuracy, even though adding new contrast mechanisms (via additional pulse sequences) can be a double-edged sword, because it increases the acquisition and evaluation times and requires additional experience and expertise to interpret the findings.

The recent exponential growth in artificial intelligence (AI) capabilities has generated new possibilities to automate vision-related tasks and support clinical decision-making and image interpretation\cite{esteva2021deep, rajpurkar2022ai}. Indeed, AI-based MRI analysis has expanded accordingly\cite{ronneberger2015u, shin2022automated, aamir2022deep, pacal2024enhancing, pacal2024novel}. However, conventional supervised AI approaches for MRI applications, such as lesion segmentation, pathology classification, and treatment response assessment, still rely on labeled data and focus on the specific anomaly used for training. As a result, their ability to generalize to unseen pathologies or clinical scenarios remains limited\cite{behrendt2026review}. Moreover, unlike natural images, which can be easily mined from the internet, medical MRI datasets are comparatively small. Other issues involve privacy concerns, institutional barriers, and data governance regulations, which often make cross-site data sharing and aggregation slow, challenging, or even infeasible\cite{willemink2020preparing}. Finally, annotating pathological MRI data is costly and time-consuming, posing a challenge for training such supervised models\cite{behrendt2026review, wang2024two}.

Recently, several large MRI databases have been collated\cite{littlejohns2020uk, zbontar2018fastmri, solomon2025fastmri}, with some leveraged to train foundation models\cite{tak2026generalizable, wu2026braindino, cox2024brainsegfounder, mei2022radimagenet}. While these models offer a route to efficient transfer learning, multi-class classification of some pathologies, and natural-language queries, they rely almost exclusively on structural (anatomical) MRI, missing the opportunity to harness the rich molecular information available through novel MRI mechanisms\cite{li2026multiplexed, cai2024non, yuan2022vivo, perlman2022quantitative}, to potentially boost their performance and broaden clinical applicability.

\noindent Unsupervised anomaly detection (UAD) is a different learning paradigm that offers an alternative strategy to address some of the above challenges\cite{nguyen2021unsupervised, meissen2022unsupervised, bercea2024diffusion, bercea2022federated}. The approach relies exclusively on data from healthy individuals for training, modeling the underlying normative data distribution, and flagging deviations as anomalies\cite{behrendt2026review}. Notably, previous work in the field focused on learning the healthy brain anatomical distribution\cite{baur2021modeling, behrendt2024patched, chen2018unsupervised}, mainly from conventional water-based structural contrast, such as T$_1$- and T$_2$-weighted images. The anomalies these frameworks can express are therefore, by construction, detected at the macroscopic level. This creates a substantial barrier for cases where the early phases of the disease are only expressed at the molecular level and do not involve visible structural alteration\cite{heim2017magnetic,jack2010hypothetical}. 

Saturation transfer (ST) MRI is a molecular imaging technique capable of detecting low-concentration metabolites and compounds that are typically invisible to conventional MRI\cite{van2011chemical}. The method uses radiofrequency (RF) pulses to selectively saturate exchangeable proton pools on the compound of interest, followed by a chemical exchange with the bulk water pool, to generate contrast related to the concentration of the targeted compound\cite{zaiss2022theory}. Critically, this may detect changes in tissue components that occur before any alterations in structural MRI. For example, in a longitudinal cuprizone multiple sclerosis (MS) mouse model, there was a significant decrease in the quantitative semisolid magnetization transfer (MT) and relayed nuclear Overhauser effect (rNOE) proton volume fractions in the corpus callosum as early as week 4 of cuprizone feeding, presenting a more sensitive detection ability compared to T$_1$ and T$_2$ changes\cite{ben2026quantitative}. Similar trends were reported in clinical settings, where changes in the semisolid MT signal preceded conventional MRI detection of MS lesions by up to two years\cite{pike2000multiple}. Similarly,  ST MRI detected glutamate changes in tauopathy up to 7 months before volume alterations\cite{crescenzi2017longitudinal}, and correctly revealed the seizure focus in drug-resistant temporal lobe epilepsy, classified as nonlesional by conventional MRI\cite{davis2015glutamate}. However, although saturation transfer MRI offers molecular insights and new clinical opportunities, it lacks the large, open, and annotated pathological databases that would be useful for automatic interpretation by classical supervised learning. Even when an unsupervised anomaly detector trained on densely sampled conventional Z-spectra was recently used to detect ischemic stroke in rats\cite{zhao2026mamba}, it required more than 20 min of acquisition time
per slice, which substantially hinders clinical translation.

Here, we propose a subject-level unsupervised anomaly detection framework trained solely on healthy-subject ST image data.  The framework is designed to address all three of the described issues by: (1) searching for anomalies using a molecular rather than an anatomical contrast; (2) operating on non-steady-state and rapid ST protocols\cite{nagar2025multi, vladimirov2025quantitative} acquired in under 43 s (for either single-slice, multi-slice, or whole-brain imaging), so the added molecular information does not substantially increase scan time; (3) offering automatic anomaly flagging, alleviating the need for expert reading of the ST images.

\section*{Results and discussion}
\subsection{Molecular UAD framework overview.}
\begin{figure}
    \centering
    \linespread{1}\fontsize{9}{11}\selectfont
    \includegraphics[width=\textwidth,height=0.5\textheight,keepaspectratio]{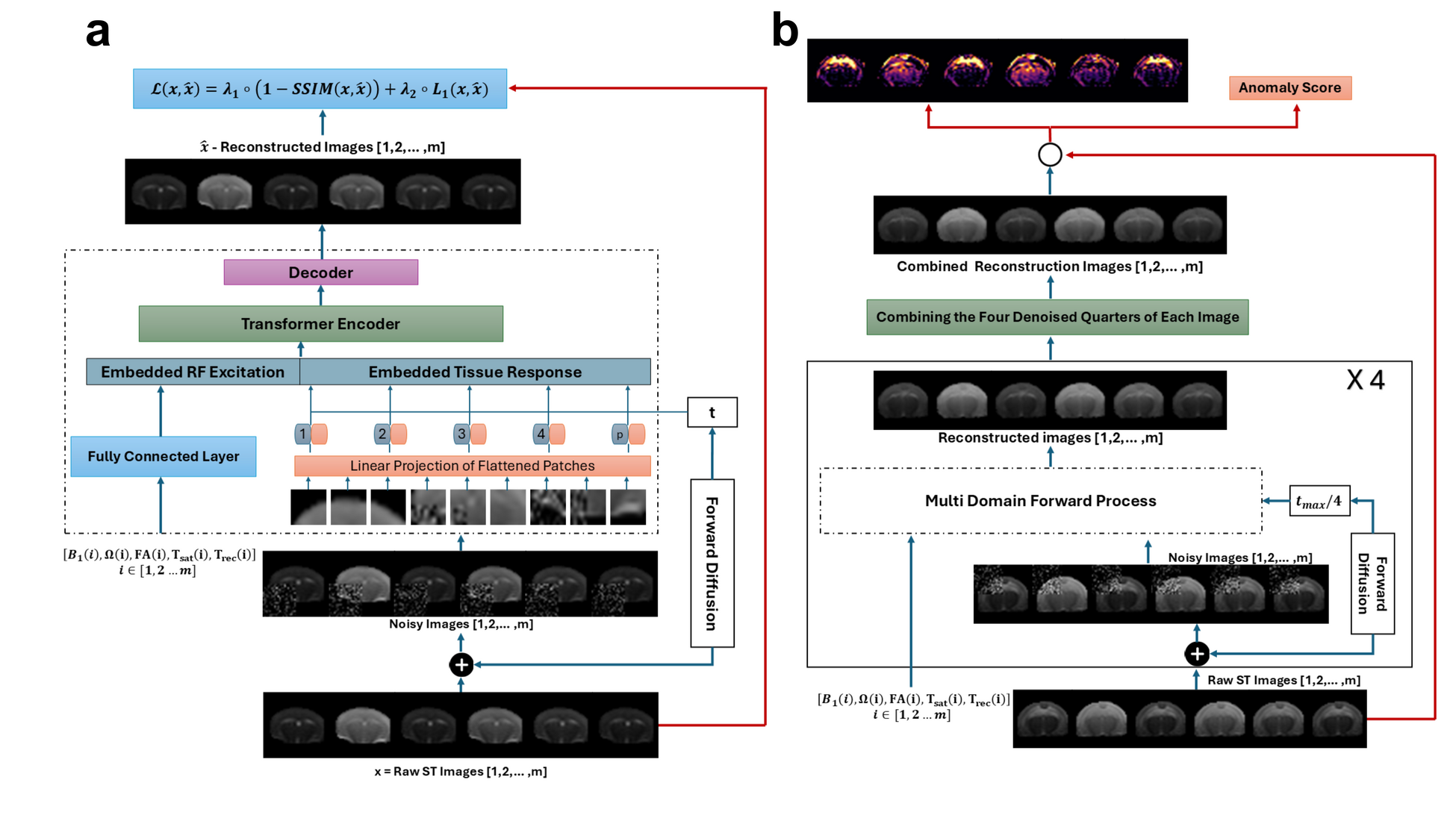}
    \caption{\textbf{Normative representation learning of molecular saturation transfer MRI.}
    \textbf{a} Training schematics. The model receives a sequence of $M=6$ saturation
    transfer (ST) images together with the radiofrequency excitation parameter tensor
    for acquisition, comprising the saturation pulse power $B_1$, frequency offset
    $\Omega$, readout flip angle FA, saturation duration $T_{\mathrm{sat}}$ and recovery
    time $T_{\mathrm{rec}}$ of each image. Forward-diffusion noise is added to one
    randomly selected spatial quarter of every image in the sequence at a randomly
    sampled timestep $t$. The corrupted sequence is tokenized by linear projection of
    flattened patches and combined with an embedding of $t$, while the excitation
    parameters are embedded separately through a fully connected layer. The two
    embeddings are concatenated, encoded by a vision transformer, and decoded by a
    convolutional decoder. This enables the uncorrupted sequence to be reconstructed
    under a combined SSIM and $L_1$ loss in which the $L_1$ term is restricted to the
    corrupted quarter.
    \textbf{b} Inference schematics. The same procedure is applied at a fixed timestep
    $t = T/4$, with each of the four quarters corrupted and reconstructed in turn and
    only the reconstructed quarter retained; the four quarters are then recombined into
    a complete normative image series. Comparing this series with the acquired input
    yields a spatial anomaly map and a single-valued anomaly score per subject and
    protocol.}
    \label{fig:architecture}
\end{figure}

The proposed framework is composed of four main elements (\textbf{Fig.~\ref{fig:architecture}}): (1) Rapid molecular ST image acquisition under non-steady-state conditions; (2) Learning a joint representation of radiofrequency excitation and the corresponding tissue response, as a physical-model-free means to encode molecular information; (3) Utilizing a vision transformer for efficient spatiotemporal representation; (4) Injecting diffusion model-based noise to facilitate normative learning\cite{pinaya2022fast}. The acquisition protocols are based on previously established quantitative ST pulse sequences\cite{vladimirov2025quantitative}, shortened to be acquired in 21-42 seconds (see Methods section), and designed to encode different molecular targets, including the semisolid MT, amide, aliphatic rNOE, and guanidinium proton pools (\textbf{Fig. S1}). The computational model takes the sequence of six ST-MRI images as input together with the RF excitation parameter tensor used to collect the data\cite{nagar2025multi}, and is trained to reconstruct the corresponding normative images (\textbf{Fig.~\ref{fig:architecture}a}). During inference, the subject-specific normative output is compared with the ST-based input to generate a spatial anomaly map and a single-valued anomaly score (\textbf{Fig.~\ref{fig:architecture}b}).

\subsection{UAD in neurological disease, aging, and healthy mouse models.}
A total of 82 mice were imaged on a 7T preclinical MRI scanner. The model was trained on wild-type
healthy mice only, but the validation and test sets also included mice with cuprizone-induced
MS, 1-methyl-4-phenyl-1,2,3,6-tetrahydropyridine (MPTP)-induced Parkinson's disease
(PD), brain cancer, and aging (\textbf{Tables S1-S2}). Of the 18 test animals, the five
MS and the five PD mice were each imaged before and after disease induction, so that their ten baseline scans formed the healthy test group, while the four brain cancer and four aging mice were imaged only once. The framework was therefore evaluated in two settings: within-subject monitoring for MS and PD, in which each animal served as its own normative reference, and subject-independent screening for brain cancer and aging. 

\begin{figure}
    \centering
    \linespread{1}\fontsize{9}{11}\selectfont
    \includegraphics[width=\textwidth,height=0.5\textheight,keepaspectratio]{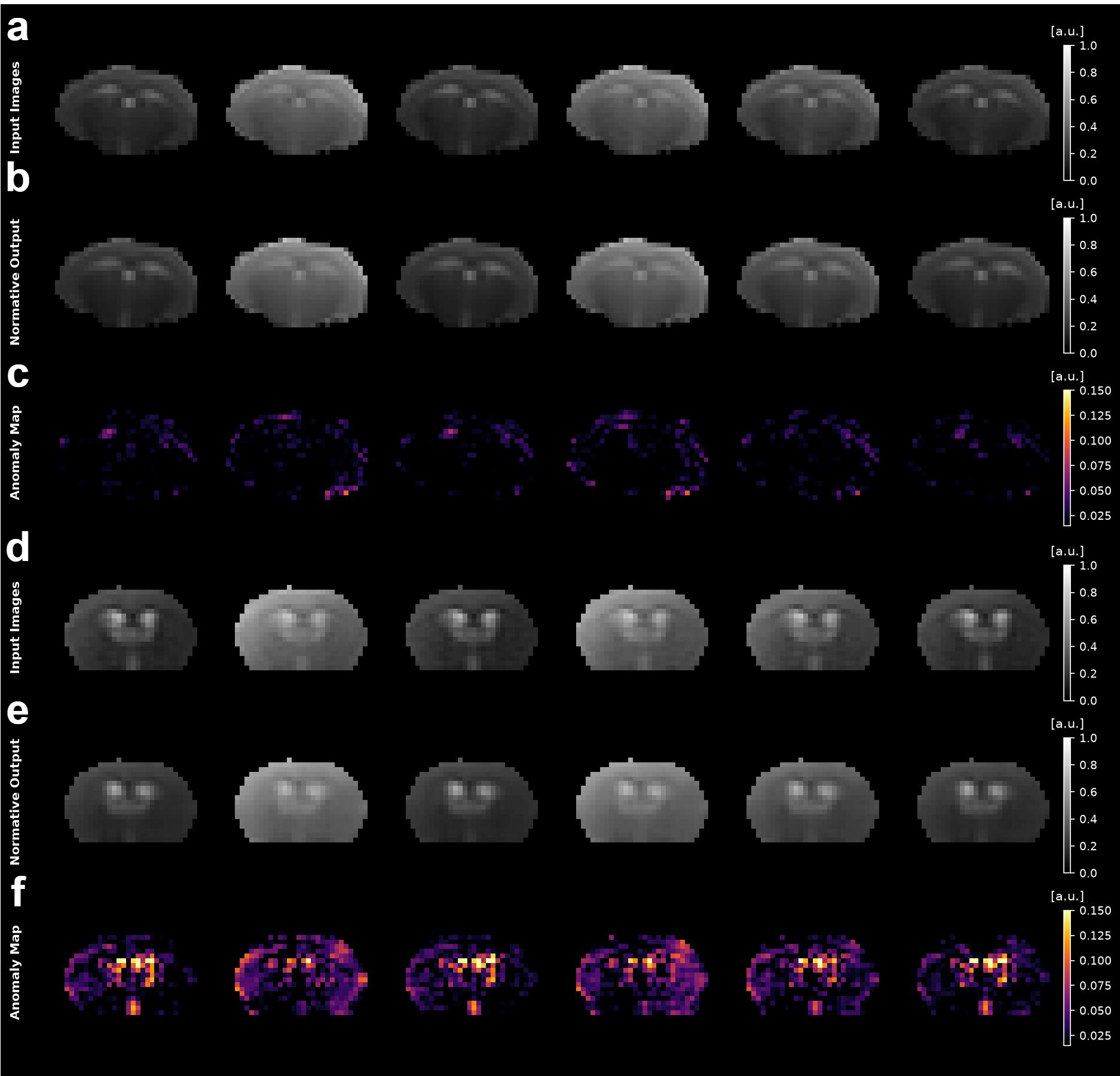}
    \caption{\textbf{Representative intermediate inference output for a healthy and an
    anomalous test mouse.} Data were acquired with the rapid semisolid magnetization
    transfer (MT) protocol and are shown for one unseen wild-type mouse (\textbf{a-c}) and
    one unseen mouse with Parkinson's disease (PD, \textbf{d-f}) from the test set.
    \textbf{a,d} The six acquired ST images that form the model input, shown left to right
    in acquisition order.
    \textbf{b,e} The corresponding normative image series reconstructed by the model.
    \textbf{c,f} Anomaly maps, computed as the voxel-wise absolute difference between the
    acquired and reconstructed images. The residual error in the wild-type mouse is low and
    sparsely distributed, but is higher and more spatially organized in the anomalous mouse.}
    \label{fig:example_mice}
\end{figure}

\begin{figure}
    \centering
    \linespread{1}\fontsize{9}{11}\selectfont
    \includegraphics[width=\textwidth,height=0.5\textheight,keepaspectratio]{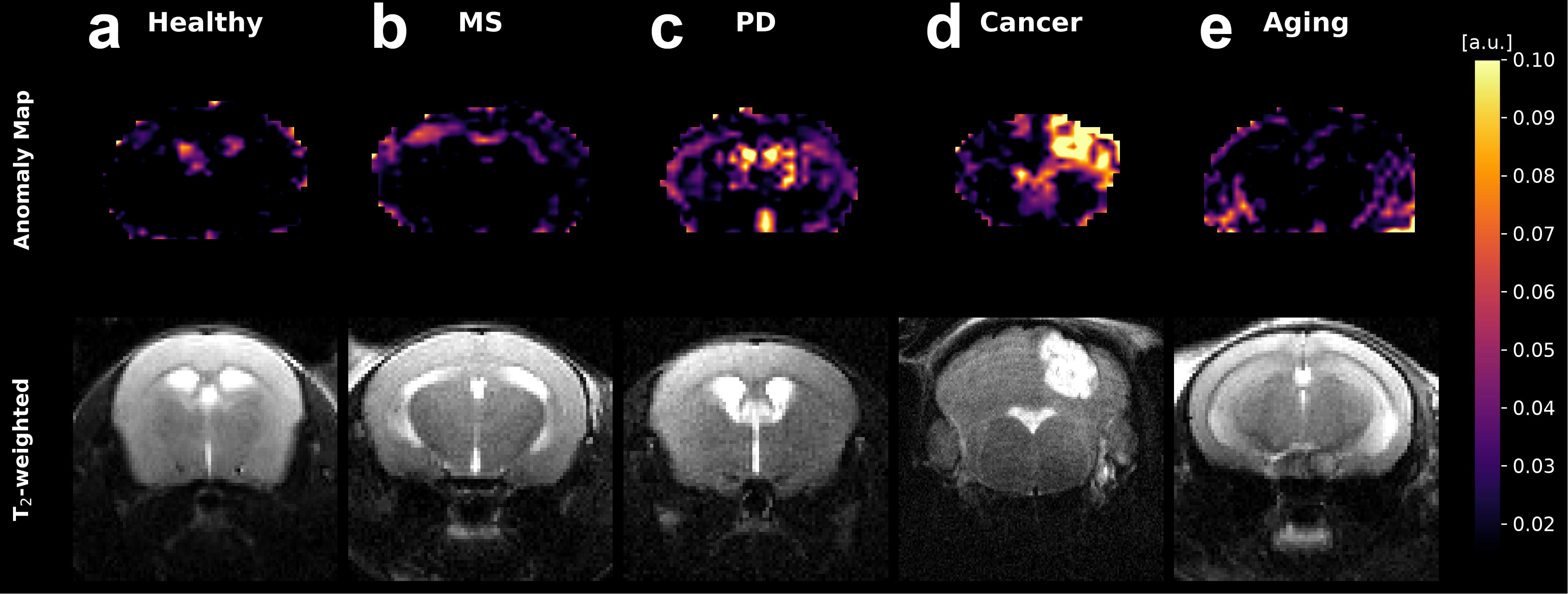}
    \caption{\textbf{Anomaly maps across five neurological categories.}
    \textbf{Top}, anomaly maps, obtained by averaging the six per-image anomaly maps of each
    animal, for representative test mice: \textbf{a} wild-type healthy; \textbf{b}
    cuprizone-induced multiple sclerosis (MS); \textbf{c} MPTP-induced Parkinson's disease
    (PD); \textbf{d} O1 cancer model; \textbf{e} 1.5-year-old aging. The wild-type map shows only sparse, low-intensity residual signal, whereas the
    anomalous animals show spatially organized elevations. 
    \textbf{Bottom}, the corresponding conventional T$_2$-weighted images of the same
    animals. The structural images are
    substantially less informative than the anomaly maps for the MS and PD cases, in which
    no structural lesion is visible.}
    \label{fig:comp_mice}
\end{figure}

Representative intermediate outputs for test-set mice comprising the acquired images, the normative reconstruction, and the resulting per-image anomaly maps, are presented in \textbf{Fig.~\ref{fig:example_mice}} and \textbf{Figs. S2-S3}. These were then averaged into a single anomaly map per animal
(\textbf{Fig.~\ref{fig:comp_mice}}). Healthy mice showed only sparse residual signal at the brain edges and ventricles (\textbf{Fig.~\ref{fig:comp_mice}a}). In contrast, there was an elevated anomaly signal in the corpus callosum in MS, the striatum in PD, the tumor location in the cancer model, and the ventricular region in aging mice (\textbf{Fig.~\ref{fig:comp_mice}b-e, respectively}). This accords with the expected phenotype of each model\cite{tagge2016spatio, ben2026quantitative, shmuely2025quantitative, bagga2016mapping,
rousso2021p32, chen2011lifespan}.

\begin{figure}
    \centering
    \linespread{1}\fontsize{9}{11}\selectfont
    \includegraphics[width=\textwidth,height=0.5\textheight,keepaspectratio]{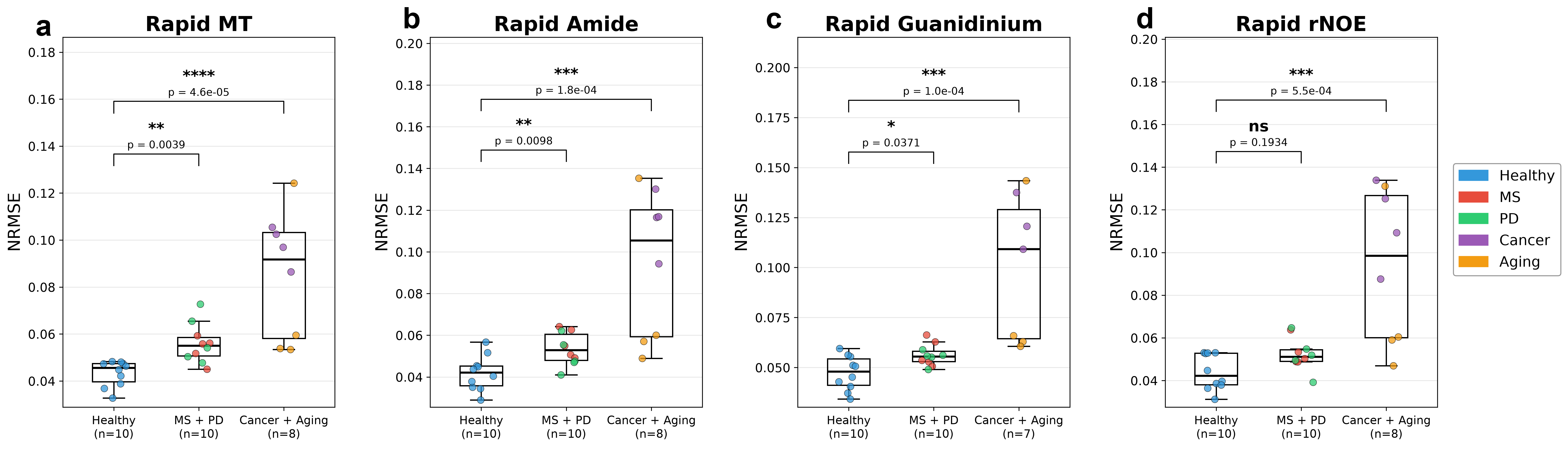}
    \caption{\textbf{Subject-level anomaly scores separate anomalous from wild-type mice
    across four molecular protocols.} Normalized root-mean-square error (NRMSE) between the
    acquired and reconstructed image series, computed per scan, for rapid MT (\textbf{a}),
    rapid amide (\textbf{b}), rapid guanidinium (\textbf{c}), and rapid rNOE (\textbf{d}).
    Each point is one scan, colored by category (healthy, MS/PD, cancer, aging); box plots
    show the median and interquartile range, and whiskers extend to the most extreme points
    within 1.5 times the interquartile range. Ten of the 18 test animals contribute both a
    pre-induction baseline scan and a post-induction scan, so the MS and PD animals were
    compared with their own baselines using an exact two-sided Wilcoxon signed-rank test
    ($n = 10$ matched pairs, both models pooled), whereas the brain cancer and aging
    animals, which have no baseline, were compared with the ten wild-type scans using an
    exact two-sided Mann--Whitney U test ($n = 10$ versus 8; $n = 10$ versus 7 for rapid
    guanidinium). Exact $p$ values are given above each bracket; ns, not significant
    ($p > 0.05$); *$p < 0.05$; **$p < 0.01$; ***$p < 0.001$; ****$p < 0.0001$.}
    \label{fig:nrmse}
\end{figure}

The single-valued per-subject anomaly scores obtained for different molecular ST target compounds are presented in \textbf{Fig.~\ref{fig:nrmse}}. The best separation between the normative and anomaly groups was obtained for rapid semisolid MT imaging (\textbf{Fig.~\ref{fig:nrmse}a}), with values of p$<$0.004 for the paired MS and PD animals and p$<$0.0001 for the cancer and aging animals, yielding an AUROC of 0.961 (95\% confidence interval (CI): 0.87-1.00) and an AUPRC of 0.982 (95\% CI: 0.94-1.00). The next best separations were given by of the rapid amide imaging protocol (p$<$0.01 for the paired MS and PD animals and p$<$0.001 for cancer and aging, AUROC=0.906, 95\% CI 0.76-1.00; AUPRC=0.946, 95\% CI 0.85-1.00), followed by guanidinium (p$<$0.04 for the paired PD and MS animals and p$<$0.001 for cancer and aging, AUROC=0.865, 95\% CI 0.71-0.98; AUPRC=0.921, 95\% CI 0.81-0.99), and rNOE (p = 0.19, not significant, for the paired MS and PD animals, and p$<$0.001 for cancer and aging, AUROC=0.833, 95\% CI 0.64-0.96; AUPRC=0.913, 95\% CI 0.80-0.98). 

With respect to pathology, all the molecular protocols studied clearly differentiated between healthy mice and those with cancer (AUROC = 1.000, AUPRC = 1.000). The best separations between aging mice and younger healthy counterparts were given by both the rapid guanidinium and the semisolid MT protocols (AUROC = 1.000, AUPRC = 1.000), followed by rapid amide (AUROC 0.950, 95\% CI 0.77-1.00; AUPRC 0.917, 95\% CI 0.62-1.00) and rNOE (AUROC 0.900, 95\% CI 0.60-1.00; AUPRC 0.875, 95\% CI 0.43-1.00). For PD, the semisolid MT protocol yielded the best performance (AUROC 0.960, 95\% CI 0.85-1.00; AUPRC 0.943, 95\% CI 0.78-1.00), ahead of rapid amide (AUROC 0.800, 95\% CI 0.50-1.00; AUPRC 0.687, 95\% CI 0.31-1.00), guanidinium (AUROC 0.740, 95\% CI 0.44-0.98; AUPRC 0.534, 95\% CI 0.25-0.98) and rNOE (AUROC 0.720, 95\% CI 0.25-1.00; AUPRC 0.677, 95\% CI 0.22-1.00). For MS, the semisolid MT and rapid amide protocols reached the same AUROC (0.900, 95\%
CI 0.62-1.00), with semisolid MT ahead on AUPRC (0.900 versus 0.826). Pooling the four protocols by averaging each animal's NRMSE across them into a single per-animal
score yielded an AUROC of 0.928 (95\% CI 0.81-1.00) and an AUPRC of 0.963 (95\% CI 0.90-1.00). For MS and PD, where each animal
provided its own pre-induction reference, the framework detected molecular deviation within the
subject, which corresponds to longitudinal monitoring of a known individual. For brain cancer and
aging, where no prior scan existed, the same model and the same learned normative distribution
separated affected animals from an independent healthy group, which is analogous to cross-sectional
screening. 

A visual demonstration of the  within animal changes is available in \textbf{Fig. S4}, which present the anomaly maps of the three MS mice at baseline and at week 7 of cuprizone feeding. At baseline, the reconstruction error was low and sparsely distributed, without a consistent spatial pattern. However, by week 7, the signal was higher and spatially organized, most prominently along the corpus callosum, the region where cuprizone-induced demyelination reduces the semisolid MT signal\cite{ben2026quantitative}. 

To evaluate the computational consistency of the framework (as the forward-diffusion noise applied at inference is stochastic), the entire test-set
evaluation was repeated for 20 independent inference seeds (\textbf{Table~\ref{tab:ours}}). The protocol ordering
was preserved throughout, with the semisolid MT protocol yielding an AUROC of 0.951 $\pm$ 0.023 and
an AUPRC of 0.975 $\pm$ 0.012 for healthy versus all anomalies, followed by rapid amide (0.910 $\pm$
0.023 and 0.948 $\pm$ 0.013), guanidinium (0.852 $\pm$ 0.022 and 0.910 $\pm$ 0.016) and rNOE (0.851
$\pm$ 0.031 and 0.921 $\pm$ 0.020); the brain cancer comparison showed complete separation in every
seed (AUROC = AUPRC = 1.000, zero variance). Among the four protocols, rapid semisolid MT yielded the best separation for all anomalies combined (\textbf{Table~\ref{tab:ours}}). We postulate that this can be attributed to the proton volume fraction being substantially larger than the other targeted pools\cite{stanisz2005t1}, which provides the highest signal-to-noise ratio (SNR) for the given short acquisition protocols used. The result is also consistent with previous quantitative studies of the same disease models\cite{ben2026quantitative, shmuely2025quantitative}, where the semisolid MT proton volume fraction in the corpus callosum or striatum of MS and PD mouse models, respectively, was most significantly altered or sensitive compared to other chemical exchange saturation transfer (CEST) compounds. Similarly, the results in the brain tumor-bearing animals accord with previous reports that quantitative MT can separate glioblastoma from normal-appearing white matter and progressive from non-progressing cases\cite{mehrabian2018quantitative}. Notably, the semisolid fraction decreased in MS but increased in PD\cite{ben2026quantitative, shmuely2025quantitative}. A reconstruction-based anomaly score is sensitive to deviations in either direction, so a single normative model can capture both.

\begin{table}
    \centering
    \linespread{1}\fontsize{9}{11}\selectfont
    \caption{\textbf{Anomaly detection performance of the proposed framework for each
    molecular protocol and anomaly class.} Area under the receiver operating characteristic
    curve (AUROC) and area under the precision-recall curve (AUPRC) for the mouse test set,
    reported as mean $\pm$ s.d. across 20 independent inference seeds of the same trained
    model. Rows are the four rapid ST protocols; the final row, ``All Protocols Averaged'',
    denotes the score obtained by averaging the NRMSE of each animal across the four
    protocols into a single per-animal value before computing the metrics. Columns are the
    individual anomaly classes and all anomaly classes combined, each compared with the ten
    wild-type test scans. Group sizes are $n = 10$ versus 4 (cancer, aging), 10 versus 5
    (PD, MS) and 10 versus 18 (all anomalies combined); for rapid guanidinium the
    corresponding cancer and combined comparisons are 10 versus 3 and 10 versus 17. Bold
    indicates the best value in each column. Cells with zero standard deviation indicate
    complete separation in every seed. Cell shading encodes the score, as given by the
    color bar.}
    \label{tab:ours}
    \includegraphics[width=\textwidth,height=0.5\textheight,keepaspectratio]{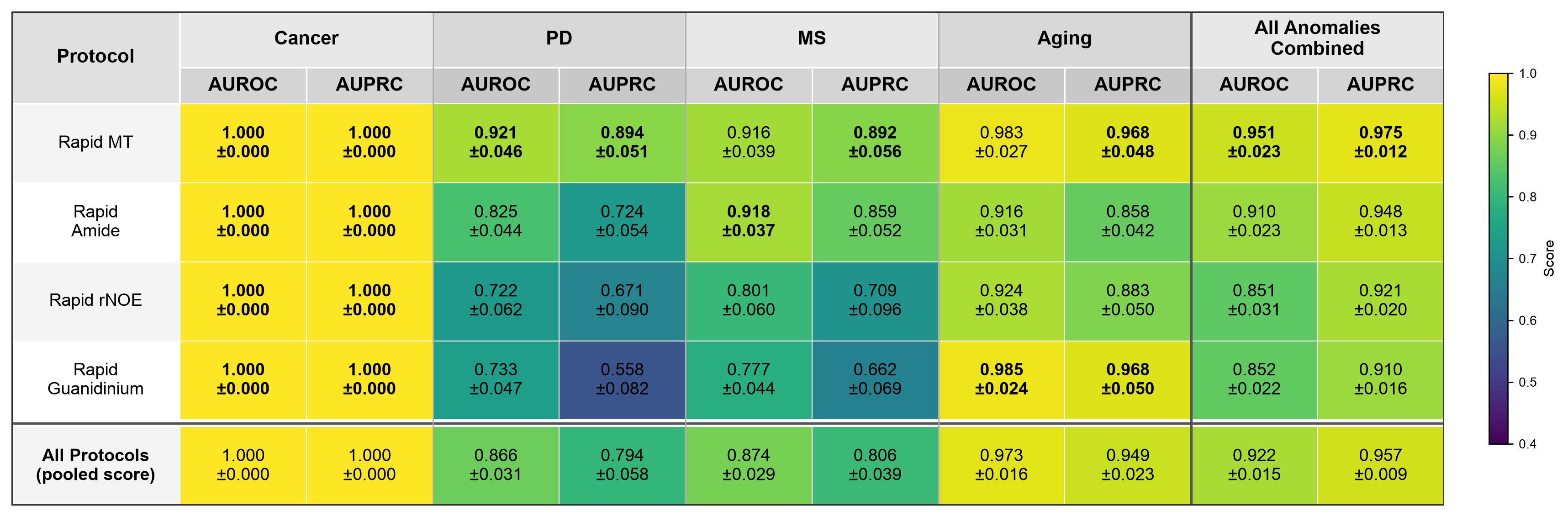}
\end{table}

This ability to detect both increases and decreases in semisolid MT highlights a broader property of normative learning. More generally, the framework does not learn disease-specific signatures, but rather a representation of healthy molecular tissue behavior. Deviations from this learned distribution are detected regardless of whether they arise from neurodegeneration, demyelination, tumor growth, or aging. This disease-agnostic formulation may be particularly valuable in situations where comprehensive labeled datasets are unavailable or where previously unseen pathologies are encountered. More broadly, the superior performance of semisolid MT suggests that anomaly detection may benefit from contrasts that probe the overall macromolecular environment rather than a single biochemical process. Semisolid MT is influenced by myelin content, tissue integrity, cellular composition, and extracellular matrix properties \cite{van2018magnetization, stanisz2005t1}, all of which can be altered in diverse pathologies. As a result, it may provide a particularly rich platform for detecting departures from normative tissue organization, independent of the specific underlying disease.

\subsection{Comparison with alternative methods.}
The proposed method was compared with three recent UAD  models: masked diffusion for brain anomaly detection (MAD-AD)\cite{beizaee2025mad}, anomaly detection with denoising diffusion probabilistic models (Ano-DDPM)\cite{wyatt2022anoddpm} and aggregated normative diffusion (ANDi)\cite{frotscher2025unsupervised}. Notably, alternative UAD approaches use human anatomical MRI data as input. For fair comparison, we retrained these methods using mouse anatomical MRI data, as described in the Methods section. To ensure that the reported performance reflects the capabilities of the method rather than a single stochastic evaluation, we evaluated each model with 20 independent inference seeds, as in the evaluation of our proposed framework. The results are presented in \textbf{Table~\ref{tab:T2}}. Across 20 independent inference seeds, the proposed framework achieved an AUROC of
0.951 $\pm$ 0.023 and an AUPRC of 0.975 $\pm$ 0.012 for healthy versus all anomalies, compared with
0.851 $\pm$ 0.010 and 0.922 $\pm$ 0.006 for MAD-AD, 0.829 $\pm$ 0.025 and 0.914 $\pm$ 0.012 for
Ano-DDPM, and 0.787 $\pm$ 0.013 and 0.894 $\pm$ 0.008 for ANDi. The proposed framework ranked first in
every seed, with a mean AUROC gain of 0.099 $\pm$ 0.025 over MAD-AD, 0.121 $\pm$ 0.033 over Ano-DDPM
and 0.164 $\pm$ 0.021 over ANDi. Brain cancer was separated perfectly by all four methods
(AUROC = AUPRC = 1.000, zero variance), whereas the proposed molecular framework demonstrated the greatest advantage in considering MS and PD pathologies.

\begin{table}
    \centering
    \linespread{1}\fontsize{9}{11}\selectfont
    \caption{\textbf{Comparison with alternative unsupervised anomaly detection methods on
    their intended anatomical input.} AUROC and AUPRC, reported as mean $\pm$ s.d. across
    the same 20 inference seeds. The proposed framework is shown operating on molecular ST
    data, using the rapid MT protocol alone and the four protocols combined, and
    additionally after retraining on the T$_2$-weighted images of the same animals. MAD-AD,
    Ano-DDPM and ANDi were developed for anatomical brain MRI and are shown here retrained
    on anatomical images. Columns, group sizes, bold and shading are as in
    Table~\ref{tab:ours}.}
    \label{tab:T2}
    \includegraphics[width=\textwidth,height=0.5\textheight,keepaspectratio]{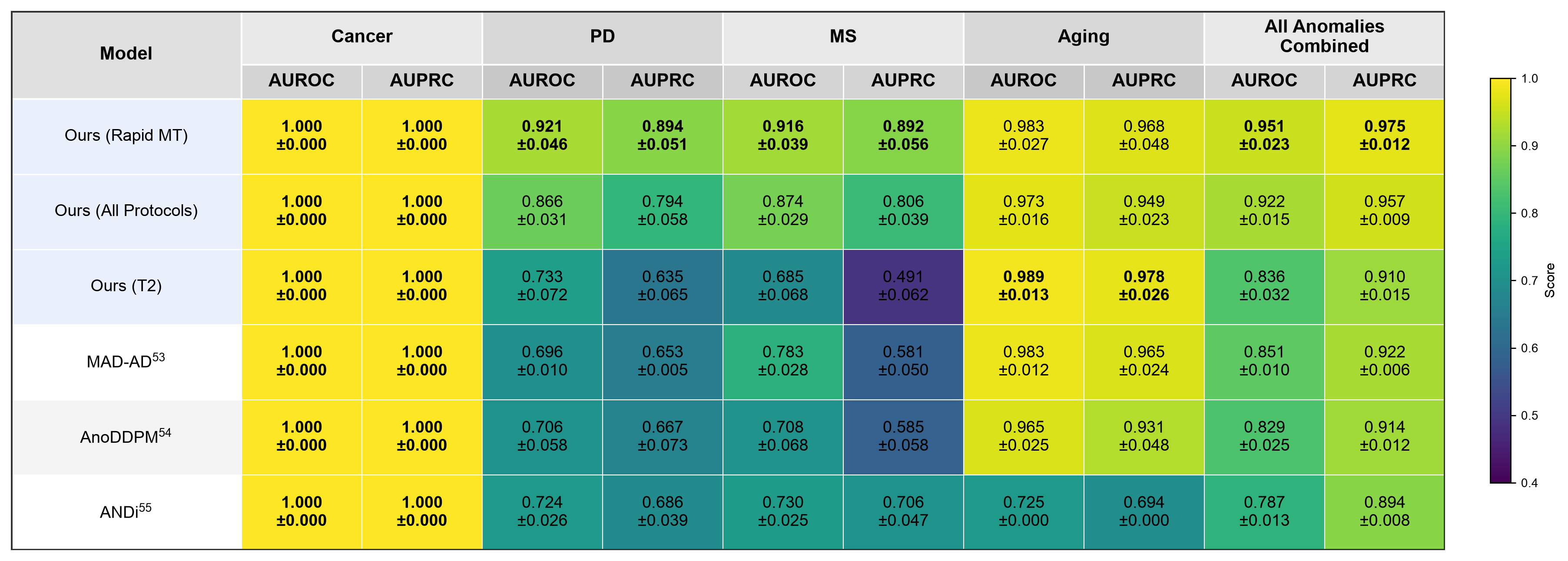}
\end{table}

\subsection{Ablation studies}

\subsection{Effect of acquisition parameter awareness and noise.}
The proposed framework is physics-aware, in that the radiofrequency excitation parameters of each acquisition are linearly projected into the same embedding space as the image patches (\textbf{Fig.~\ref{fig:architecture}a}). At the same time, normative learning is enforced during training by injecting forward-diffusion noise into one randomly selected quarter of each input image series, which the network must later restore. To evaluate the separate and joint effects of these two methodological components, we trained the model from scratch with and without the noise addition strategy (across the full image, a quarter of the image, or none), and with and without acquisition parameter awareness (\textbf{Table S3}).

The results indicate that noise injection was a dominant factor in performance, because removal reduced the all-anomaly AUROC under the rapid MT protocol from 0.951 $\pm$ 0.023 to 0.767 $\pm$ 0.000, and every noise-based configuration matched or exceeded the noise-free configurations in every anomaly class. Restricting the noise to a quarter of the image rather than the full image was the second largest contributor (0.951 $\pm$ 0.023 versus 0.866 $\pm$ 0.035), and acquisition parameter awareness added a further 0.032 (0.951 $\pm$ 0.023 versus 0.919 $\pm$ 0.021). Quarter-image noise addition combined with the excitation parameters, as used in our proposed model, achieved the best performance for the all-anomaly, MS and PD comparisons under both the rapid MT and the all-protocols datasets, and brain cancer was separated perfectly by every configuration. The single exception was aging, for which the configuration without parameter awareness performed marginally better under rapid MT (0.985 $\pm$ 0.024 versus 0.983 $\pm$ 0.027) and reached perfect separation when using NRMSE averaged from all protocols (1.000 $\pm$ 0.000 versus 0.973 $\pm$ 0.016). Accordingly, parameter awareness contributed most to the molecularly expressed pathologies (AUROC gains of 0.091 for MS and 0.027 for PD under rapid MT). 

Consistent with previous reports of the value of explicit RF excitation information to ST-MRI, removing parameter awareness reduced the all-anomaly AUROC from 0.922 $\pm$ 0.015 to 0.883 $\pm$ 0.015 and the AUPRC from 0.957 $\pm$ 0.008 to 0.941 $\pm$ 0.008, in all protocols \cite{nagar2025multi}. These results highlight the importance of acquisition-aware learning for molecular MRI analysis. The contrast in ST-MRI arises from the interaction between tissue properties and the applied radiofrequency (RF) excitation scheme. Therefore, explicitly conditioning the model on the acquisition parameters enables the network to interpret tissue responses within the \textit{temporal} (spin-history) and \textit{biophysics} (RF-to-tissue response) context in which they were generated. The advantage of quarter-image over full-image noise suggests that the un-noised regions provide \textit{spatial} signal context that guides the normative reconstruction\cite{nguyen2021unsupervised, beizaee2025mad, behrendt2024patched}. 

\subsection{Effect of architecture vs. molecular data input.} To examine whether the superior performance of our proposed framework is due to the molecular image input or the AI architecture, we retrained our method on the same anatomical T$_2$-weighted data used by the alternative approaches (\textbf{Table~\ref{tab:T2}}). Performance on the brain cancer cohort remained optimal (AUROC = AUPRC = 1.000), and the anatomically expressed ventricular enlargement of aging was detected marginally better than with molecular input (AUROC
0.989 $\pm$ 0.013 versus 0.983 $\pm$ 0.027). However, the performance on PD and MS, was reduced from 0.916 $\pm$ 0.039 to 0.685 $\pm$ 0.068 for MS and from 0.921 $\pm$ 0.046 to
0.733 $\pm$ 0.072 for PD. The values for all combined anomalies were reduced from 0.951 $\pm$ 0.023 to 0.836 $\pm$ 0.032, suggesting that it is the molecular information, rather than strictly the architecture alone, that drives the detection of these pathologies.

As a corollary, we then retrained the alternative methods to receive the same molecular ST data used in our pipeline (without the excitation parameters, which their architectures do not take as input; \textbf{Table~\ref{tab:ST}}), across 20 inference seeds. Each method was evaluated both on the best-performing rapid MT protocol alone and on all four protocols, with the per-animal NRMSE values averaged across protocols into a single score. Under the rapid MT protocol, MAD-AD reached an AUROC of 0.898 $\pm$ 0.018 and an AUPRC of 0.947 $\pm$ 0.011, followed by Ano-DDPM (0.787 $\pm$ 0.040 and 0.890 $\pm$ 0.018) and ANDi (0.676 $\pm$ 0.010 and 0.847 $\pm$ 0.003); with the four protocols averaged, the corresponding AUROC values were 0.811 $\pm$ 0.009, 0.718 $\pm$ 0.050 and
0.800 $\pm$ 0.008. Notably, for all anomalies combined, only MAD-AD benefited from the molecular input relative to its native anatomical input (0.898 versus 0.851), whereas Ano-DDPM and ANDi performed worse (0.787 versus 0.829 and 0.676 versus 0.787), indicating that molecular ST data are not exploited equally by every UAD architecture.

\begin{table}
    \centering
    \linespread{1}\fontsize{9}{11}\selectfont
    \caption{\textbf{Comparison with alternative methods given the same molecular input.}
    AUROC and AUPRC, reported as mean $\pm$ s.d. across the same 20 inference seeds, for
    the proposed framework and for MAD-AD, Ano-DDPM and ANDi after retraining on the
    molecular ST data. Each method is shown on the rapid MT protocol alone and on the four protocols
    combined. Columns, group sizes, bold and shading are as in Table~\ref{tab:ours}.}
    \label{tab:ST}
    \includegraphics[width=\textwidth,height=0.5\textheight,keepaspectratio]{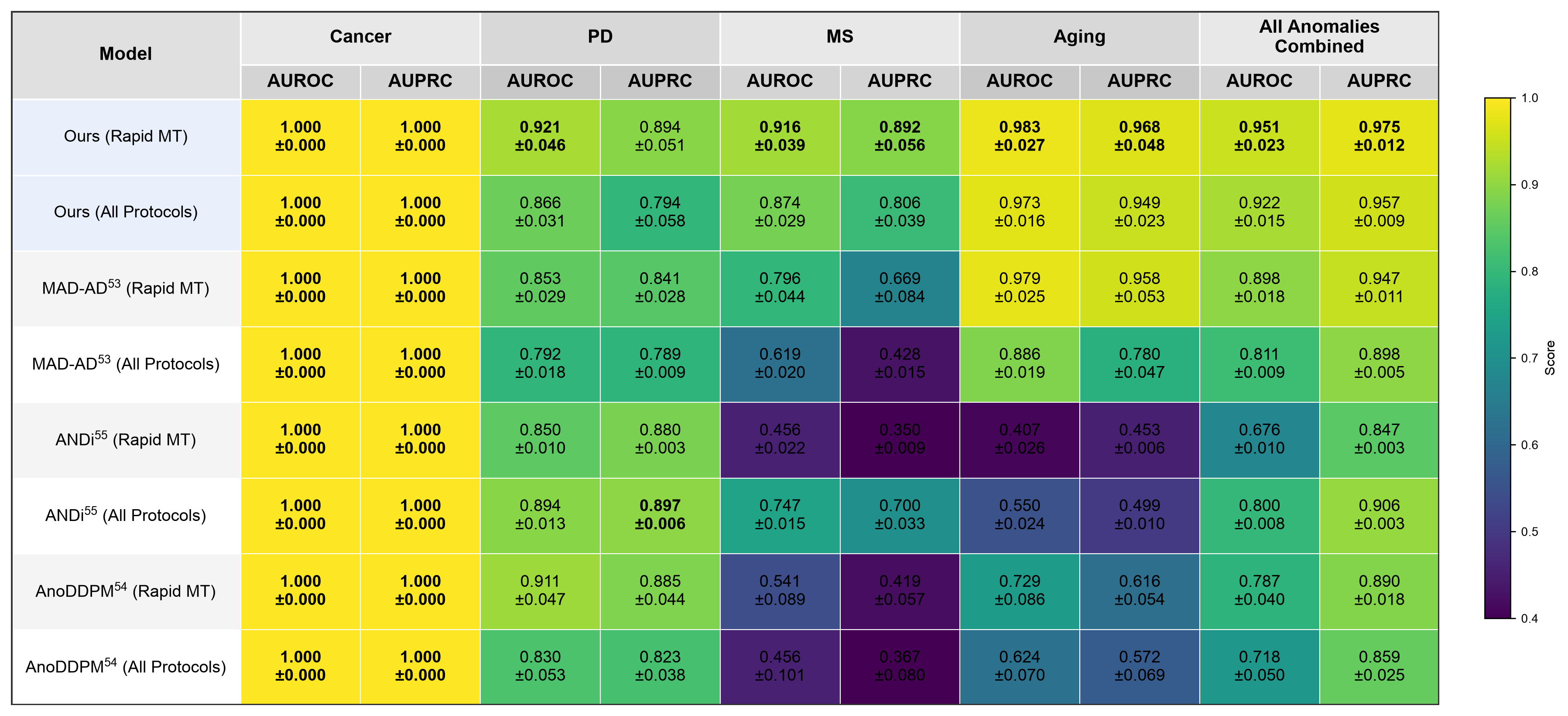}
\end{table}

At the pathology level (\textbf{Tables~\ref{tab:T2}-\ref{tab:ST}}, rapid MT), switching from anatomical to molecular input improved PD detection for all four methods (AUROC, proposed framework 0.733 to 0.921, MAD-AD 0.696 to 0.853, Ano-DDPM 0.706 to 0.911 and ANDi 0.724 to 0.850). In contrast, MS detection improved substantially only for the proposed framework (0.685 to 0.916) and marginally for MAD-AD (0.783 to 0.796), with decreases in Ano-DDPM and ANDi for both MS (0.708 to 0.541 and 0.730 to 0.456) and aging (0.965 to 0.729 and 0.725 to 0.408).

The results indicate that given the same molecular input, our proposed framework outperforms the strongest alternative, MAD-AD, in all 20 seeds, by 0.053 $\pm$ 0.014 in AUROC. However, this was not the case for anatomical input: our method retrained on $T_2$-weighted data exceeded MAD-AD in only 9 of the 20 seeds for AUROC and 4 of 20 for AUPRC, while surpassing Ano-DDPM in 12 of 20 (11 of 20 for AUPRC) and ANDi in 17 of 20 (17 of 20 for AUPRC). Taken together, we can not attribute the superiority of the proposed framework to either the architecture or the input alone, but rather to an architecture that conditions on the excitation parameters and can thereby exploit the molecular contrast. This is consistent with the loss in performance when parameter awareness was removed (\textbf{Table S3}).

\subsection{Preliminary clinical translation study.} 
As a preliminary proof of concept for clinical translation, a separate model was trained from scratch
on human data, following the same procedure as in the mouse study and using the rapid MT protocol,
which performed best there. Thirty-six participants were scanned at four sites (Tel Aviv University, TAU; Erlangen, ERL; Meir Medical Center, MMC; and Tel Aviv Sourasky University Medical Center, TASMC) on three 3T scanner models, with the acquisition harmonized across sites by a single vendor-independent pulseq-CEST
sequence definition file\cite{layton2017pulseq, herz2021pulseq}. All twelve training participants
were healthy and were scanned at a single site (TAU), and architectural changes were limited to those required by the larger human field of view and image matrix (see \textbf{Methods} section). Of the
remaining participants, five were used for validation and 19 for testing: seven healthy participants scanned at four sites, six patients with brain cancer scanned at three sites, and six patients with Alzheimer's
disease scanned at a single site (\textbf{Tables S4-S5}).

\begin{figure}
    \centering
    \linespread{1}\fontsize{9}{11}\selectfont
    \includegraphics[width=\textwidth,height=0.5\textheight,keepaspectratio]{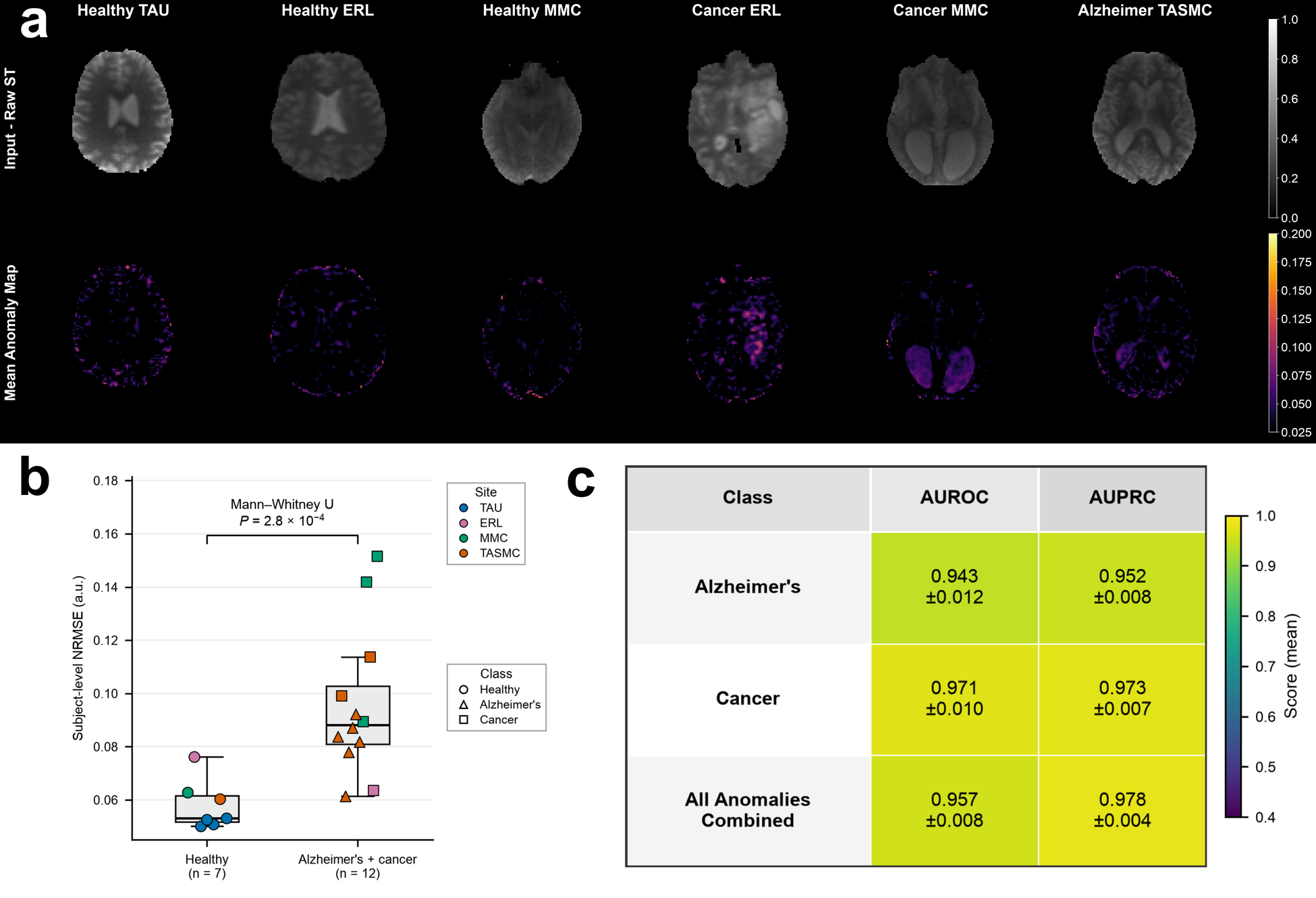}
    \caption{\textbf{Preliminary clinical translation across four imaging sites.}
    \textbf{a} First input image of the six-image sequence \textbf{(top)} and the corresponding
    anomaly map \textbf{(bottom)}, for six representative test participants: healthy participants
    from TAU, ERL, and MMC, brain cancer patients from ERL and MMC, and a patient with
    Alzheimer's disease from TASMC.
    \textbf{b} Subject-level NRMSE for all 19 test participants, with each participant's
    slice-wise scores averaged into a single value; color denotes the acquisition site and
    shape denotes the class. Box plots show the median and interquartile range, with
    whiskers extending to the most extreme points within 1.5 times the interquartile range;
    all individual participants are shown, including those beyond the whiskers. $p$, exact
    two-sided Mann--Whitney U test, healthy versus all patients.
    \textbf{c} Subject-level AUROC and AUPRC for each pathology against the healthy group,
    and for both pathologies combined against the healthy group, reported as mean $\pm$
    s.d. across 20 inference seeds. Healthy, $n = 7$; Alzheimer's disease, $n = 6$; brain
    cancer, $n = 6$. TAU, Tel Aviv University; ERL, Erlangen; MMC, Meir Medical Center;
    TASMC, Tel Aviv Sourasky University Medical Center.}
    \label{fig:human}
\end{figure}

Healthy participants showed low anomaly-map values, mostly at the brain edges, whereas patients showed elevated values (\textbf{Fig.~\ref{fig:human}a}). For the representative inference seed shown in \textbf{Fig.~\ref{fig:human}}, the framework separated healthy participants from all patients with an AUROC of 0.964 (95\% CI 0.86-1.00) and an AUPRC of 0.981 (95\% CI 0.93-1.00). The AUROC of patients with Alzheimer's disease, was 0.952 (95\% CI 0.81-1.00) and the AUPRC was 0.958 (95\% CI 0.84-1.00). For brain cancer, the AUROC was 0.976 (95\% CI 0.86-1.00) with an AUPRC of 0.976 (95\% CI 0.88-1.00). 
The all anomalies combined results for 20 seeds inference repetitions, were AUROC of 0.957 $\pm$0.008 and AUPRC of 0.978 $\pm$ 0.004. For Alzheimer’s disease, the AUROC was 0.943 $\pm$ 0.012 and the AUPRC 0.952 $\pm$ 0.008. For brain cancer, the AUROC was 0.971 $\pm$ 0.010 and AUPRC of 0.973 $\pm$ 0.007.

Within the external sites, the MMC and TASMC healthy participants scored lower than the brain cancer patients scanned there (MMC 0.063 versus 0.089-0.151; TASMC 0.060 versus 0.099-0.114; \textbf{Fig.~\ref{fig:human}b}), whereas the ERL healthy participant scored higher than the ERL cancer patient (0.076 versus 0.063). In addition, the TASMC healthy participant, whose age (65 years) lies within the range of the patients with Alzheimer’s disease scanned at that site (56-77 years), scored lower than all six of these patients. The two older healthy participants (MMC, 74 years; TASMC, 65 years) scored higher than the TAU healthy participants (42 years or younger), but lower than all but one patient with Alzheimer’s disease. 


  \subsection{Limitations and future work} 
Although the framework produces spatial anomaly maps (\textbf{Fig.~\ref{fig:comp_mice}} and \textbf{Fig.~\ref{fig:human}a}), it was designed and evaluated as a subject-level detector, and the pixel-wise capabilities remain limited. Although the maps localized to the expected regions in MS, cancer and aging (\textbf{Fig.~\ref{fig:comp_mice}}), this agreement was assessed only visually: no voxel-wise ground truth was available, so localization accuracy (e.g., Dice or pixel-level AUROC) could not be quantified. Additional factors that limit the maps concern the observation that residual reconstruction error appears in healthy subjects at the brain edges and ventricles (\textbf{Fig.~\ref{fig:comp_mice}a} and \textbf{Fig.~\ref{fig:human}a}). The mouse input images were single 2D slices, which restricts the extent of spatial details available. In humans, the subject-level score averages the error over the multiple slices acquired, so a small focal lesion contributes less than a widespread change. The maps should therefore be regarded as an aid to interpretation, indicating where the model's normative reconstruction failed, rather than a readily available lesion segmentation. 

Training a UAD framework typically requires a large amount of healthy data\cite{baur2021modeling}, whereas the present study used 48 wild-type mice and 12 healthy human participants for training. We mitigated this in two ways. In mice, each animal was scanned with various ST protocols, each acquired with a different set of excitation parameters, and every acquisition was divided into overlapping six-image sequences and augmented by rotation, yielding over 100,000 training samples (\textbf{Table S2}). Because the excitation parameters are an explicit input, these acquisitions act as complementary views of the same normative tissue rather than as redundant copies. In humans, each 3D volume was split into axial, sagittal, and coronal slices, yielding a total of 423,000 training sets (\textbf{Table S5}). Future work should increase the animal and human sample size for both training and test purposes, while adding subjects with PD and MS (and ideally additional neurological disorders) to the human test set, which currently serves as a proof of concept. 

Although preliminary, the human results suggest that the proposed approach can generalize beyond the acquisition site used for training. The main exception was the ERL healthy participant, who scored higher than the ERL cancer patient and one patient with Alzheimer's disease (\textbf{Fig.~\ref{fig:human}b}). With a single healthy participant per external site, we cannot be sure whether this reflects individual variability or site-related effects. Nevertheless, the separation of healthy participants from patients across multiple institutions and scanner models suggests that rapid molecular MRI may serve as a practical route for harmonized normative imaging studies. Larger and more diverse healthy cohorts, spanning multiple sites and a wider age range, will be required to determine the extent to which the learned healthy distribution can accommodate demographic, biological, and scanner-related variability.

The acquisition-aware paradigm introduced here could be generalized beyond ST MRI to other quantitative and molecular imaging techniques, providing a broader framework for physics-informed normative imaging in precision medicine.

\section*{Conclusions}
In summary, we developed a normative learning framework for unsupervised anomaly detection that learns the healthy brain from rapid saturation transfer MRI together with the RF excitation parameters used to acquire it. Trained on healthy subjects only, it detected MS, PD, brain cancer, and aging in mice from a single rapid semisolid MT acquisition, with MR data acquired in under 43 s and inference completed in under 6 s. The molecular input was key to detecting MS and PD, which lack visible structural alterations. Conditioning on the excitation parameters further improved detection and allowed the framework to exploit the molecular contrast better than alternative architectures given the same data. Because only healthy data and subject-level labels are required, the approach avoids the annotation burden and the scarcity of pathological data that limit supervised analysis of molecular MRI, and could support automatic flagging and prioritization of urgent cases at a negligible cost in scan time. A preliminary study at four clinical sites suggests that the framework transfers to human scanners for brain cancer and Alzheimer's disease. Future validation in larger multi-center cohorts will be required to establish clinical utility and generalizability.

\begin{methods}

\subsection{Mouse cohort}
All experiments were approved by the Tel Aviv University Institutional Animal Care and Use Committee (IACUC; protocols  TAU-MD-IL-2507-135-5, TAU-MD-IL-2309-160-5, TAU-MD-IL-2405-129-5, TAU-MD-IL-2507-137-3) and adhered to the ethical principles of the Israel National Research Council (NRC). A total of 82 mice were included in the study, comprising healthy wild-type mice C57BL/6J, male, aged 8-12 weeks and four disease/physiological models were included in the study: cuprizone-induced multiple sclerosis (MS; C57BL/6J, male; scanned at 8-20 weeks of age); MPTP-induced Parkinson's disease (PD; C57BL/6J, male, 12 weeks of age); brain cancer (WT C57BL/6 female mice injected with O1 cells and treated with UT/P32 CAR-T, 12-15 weeks of age)\cite{rousso2021p32}; and an aging model (C57BL/6J, male, 1.5 year old). The MS and PD mice were drawn from two previously described longitudinal studies of male C57BL/6J mice (Harlan, Israel)\cite{ben2026quantitative, shmuely2025quantitative}. MS mice were scanned at baseline and then at week 7 of cuprizone feeding, corresponding to peak demyelination\cite{ben2026quantitative}. PD mice were scanned at baseline, treated with four intraperitoneal (IP) injections of MPTP-HCl (14-20 mg/kg, dissolved in saline) administered every 2 h over an 8 h period (acute dosing paradigm), and rescanned 7 days after treatment\cite{shmuely2025quantitative}. 

Mice were assigned to non-overlapping training, validation, and test sets of 48, 16, and 18 animals, respectively (58.5\%, 19.5\%, and 22.0\%; \textbf{Table S1}). The training set consisted exclusively of healthy wild-type mice. The validation set included nine healthy wild-type mice, three of which subsequently underwent cuprizone induction and four MPTP induction, as well as one aging and six mice with brain cancer. The test set included five MS and five PD mice, each scanned before disease induction (baseline) and after disease establishment (week 7 of cuprizone feeding and 7 days after MPTP administration, respectively), as well as four brain cancer and four aging mice, each scanned once. The ten baseline scans from the MS and PD cohorts served as the healthy test group. Consequently, anomaly detection in the MS and PD cohorts was evaluated longitudinally by comparing post-induction scans with subject-matched baseline scans, whereas brain cancer and aging mice were assessed relative to independent healthy control animals, reflecting a subject-independent screening scenario.

\subsection{Human cohort.} Thirty-six participants, comprising healthy volunteers, patients with Alzheimer's disease, and patients with brain tumors, were enrolled across four sites:
Tel Aviv University (TAU, Israel), Erlangen (ERL, Germany); Tel Aviv Sourasky University Medical Center (TASMC, Israel); and Meir Medical Center (MMC, Israel). All studies were approved by the respective local ethics committees, and written informed consent was obtained from all participants (\textbf{Table S5}). The MMC cohort comprised four participants (two males and two females, age 49 $\pm$ 22 years, range 23-74 years), including one healthy volunteer and three patients with brain tumors classified according to the WHO Classification of Tumors of the Central Nervous System\cite{louis20212021} as Astrocytoma, IDH-mutant, WHO Grade 2; supratentorial ependymoma, ZFTA-fusion positive, WHO Grade 3; and glioblastoma, IDH-wildtype, WHO Grade 4. The ERL cohort comprised a 38-year-old male subject with glioblastoma (WHO grade IV, IDH mutation, and MGMT-promoter methylated) and one healthy 26-year-old female volunteer. The TAU cohort comprised 20 healthy participants (10 males, 10 females; age 27 $\pm$ 6 years, range 22–42 years). The TASMC cohort comprised seven clinically diagnosed patients with Alzheimer's disease (4 males, 3 females; age 70 ± 7 years, range 56–77 years), two female patients with brain metastases (aged 57 and 71 years), one from renal cell carcinoma (RCC) and one, in the left parietal lobe, from non-small cell lung cancer (NSCLC), and a healthy control subject (65-year-old male). Participants were assigned to non-overlapping training, validation, and test sets, as described in \textbf{Tables S4 and S5}.

\subsection{Preclinical MRI acquisition.}
All mouse imaging experiments were performed on a 7T MRI scanner (Bruker, Germany). During imaging, animals were anesthetized with 0.5-2\% isoflurane and continuously monitored using a physiological monitoring system (SA Instruments, NY, USA). The imaging cradle included a hot water circulation system that maintained body temperature at 37\,$^\circ$C. 
Various rapid saturation transfer protocols were applied to encode the semisolid MT, amide, aliphatic rNOE, and guanidinium proton pools into distinct signal trajectories (\textbf{Figure S1}, with six sequence variants used for training only and all validation and test animals scanned with the same four rapid protocols). The protocols were derived from previously established quantitative ST magnetic resonance fingerprinting (MRF) sequences\cite{vladimirov2025quantitative, perlman2022quantitative}, acquired within 105-210 s per full 30-image protocol. For subsequent computational analysis, only a shortened subset of six consecutive images was used as input, corresponding to an effective acquisition time of 21-42 s. The saturation pulse power was varied
between 0 and 6\,$\mu$T and the frequency offset was varied between 6 and 14 ppm for semisolid MT encoding\cite{perlman2022quantitative} or fixed at the chemical shift of the other metabolite of interest mentioned above. The flip angle was set to 60 or 90 degrees, the saturation time to 2.5 s, 3 s, or 3.5 s, and the recovery time to 1 s, 3.5 s, or 4 s (see detailed properties in \textbf{Figure S1}). A spin-echo echo-planar imaging (SE-EPI) readout was used in all cases, with a TE $=20$ ms. The field of view was $19 \times 19$\,mm$^2$ with a slice thickness of 1.5 mm and an in-plane resolution of 297\,$\mu$m, yielding a $64 \times 64$ acquisition matrix. Comparison with alternative anomaly detection methods that build on anatomical MRI as input, was enabled by acquiring multi-echo spin-echo images using the same field of view, slice thickness, and spatial resolution as the molecular imaging protocols, with TR=2000 ms and TE=30 and 75 ms. In addition, a T$_2$-weighted image with twice the in-plane resolution was also acquired for visualization purposes.

\subsection{Human MRI acquisition.}
 Imaging was performed on three 3T MRI scanners (Siemens Healthineers): a Prisma scanner (used at TAU and ERL); a Vida scanner (at MMC); and a Cima scanner (at TASMC).
All participants were scanned using the rapid MT protocol, employing the same saturation pulse powers and frequency offsets used in the preclinical study, while incorporating only the modifications necessary for implementation on clinical 3T systems. To ensure protocol harmonization across sites and scanner platforms, all acquisitions were performed using a single pulseq-CEST sequence file (.seq). 
Specifically, a spin lock saturation train (13 $\times$ 100 ms, 50\% duty-cycle), with saturation pulse power varied between 0 and 4 $\mu$T (\textbf{Fig. S1}), was used to generate 30 contrast-weighted images (later separated into different groups of six consecutive images for analysis). The saturation pulse frequency offset was varied between 6 and 14 ppm for semisolid MT encoding\cite{perlman2022quantitative}. The saturation block was fused with a 3D centric reordered snapshot EPI readout at TAU and ERL, and with a 3D snapshot GRE readout at TASMC and MMC\cite{mueller2020whole, deshmane20193d}. The resolution was 1.8 mm isotropic across a whole-
brain field of view (TAU, ERL) or 2.0 × 2.0 × 5.0 mm$^3$, 3D axial slab of 12-14 slices (TASMC and
MMC). The scan time for the six-image 3D data acquisition across all sites was 28.2-30 seconds.

\subsection{Preclinical data preprocessing.}
A series of 30 two-dimensional $64 \times 64$ images was acquired for each mouse and imaging protocol. A brain mask was drawn manually for each protocol of each mouse and applied element-wise to the image series, and a $40 \times 40$ bounding box centered on the center of mass of the first masked image was applied to all 30 images, yielding a $30 \times 40 \times 40$ array. Each array was paired with the five excitation parameters of the corresponding 30 images ($B_1$, $\Omega$, FA, $T_{\mathrm{sat}}$ and $T_{\mathrm{rec}}$), giving a $5 \times 30$ parameter array. The series was then divided into all possible sets of six consecutively acquired images, each comprising a $6 \times 40 \times 40$ image sequence and the paired $5 \times 6$ parameter tensor, which serve as separate inputs to the network. This process served two purposes: (1) to teach the network the relation between different combinations of excitation parameters and the resulting tissue response reflected in the images\cite{nagar2025multi}, and (2) to provide the first stage of training-set data augmentation. Notably, the magnetization state at the first image of each set, which depends on the saturation and readout history of all preceding images, is not provided to the network (the parameter tensor describes only the six images within the set). Because the sets start at different points in the series, each begins from a different and unknown initial condition, determined by the incomplete magnetization recovery that follows the preceding images. A vision transformer with RF excitation embeddings has previously been shown to cope with such unknown initial conditions, and satisfactory reconstructions were obtained in such cases\cite{nagar2025multi}. Training on sets drawn from all positions of the series therefore exposes the network to a range of spin histories, so that the learned normative tissue response does not depend on a single initial condition. An 8-fold rotation-based augmentation was then applied to the training set only, constituting the second stage of augmentation. The resulting numbers of training, validation, and test samples are given in \textbf{Table S2}; the test count reflects a single six-image sample per scan and protocol, less one guanidinium scan that was not acquired for one brain cancer mouse.

\subsection{Human data preprocessing.}
All images were motion-corrected and registered using Elastix\cite{klein2009elastix}. Skull removal was performed using statistical parameter mapping (SPM)\cite{ashburner2005unified}. Image slices were zero-padded to 128$\times$128 pixels, which accommodates the largest brain cross-section
under rotation and is divisible by the vision transformer patch size. Each slice was then divided into samples of six consecutive saturation scans (128$\times$128$\times$6), paired with a 2$\times$6 tensor of the
corresponding saturation pulse powers and frequency offsets (as the human study was focused on a single rapid MT protocol, where these two parameters were varied). Training set augmentation was obtained using in-plane rotation with eight different angles (\textbf{Table S5}).

\subsection{Molecular ST UAD computational framework.}
The framework learns to reconstruct the normative ST image series of a subject from a partially noise-corrupted version of that series, conditioned on the RF excitation parameters used for the acquisition (\textbf{Fig.~\ref{fig:architecture}a}). It is built on a physics-aware vision transformer\cite{nagar2025multi, dosovitskiy2020image} and comprises a training process (\textbf{Fig.~\ref{fig:architecture}a}) and an inference process
(\textbf{Fig.~\ref{fig:architecture}b}).

\noindent\textbf{Inputs.} Each sample consists of a sequence of $M=6$ consecutive ST images
$\boldsymbol{x} \in \mathbb{R}^{6 \times 40 \times 40}$ and the excitation parameter tensor
$\boldsymbol{p} \in \mathbb{R}^{5 \times M}$ used to play them out, whose five rows are the
saturation pulse power $B_1(i)$, the saturation frequency offset $\Omega(i)$, the readout flip angle
$\mathrm{FA}(i)$, the saturation duration $T_{\mathrm{sat}}(i)$ and the recovery time
$T_{\mathrm{rec}}(i)$, for $i \in [1,\dots,M]$ (\textbf{Fig. S1}).

\noindent\textbf{Forward diffusion.} During training, a timestep $t \sim \mathcal{U}(5, 500)$ was
sampled from the range $t \in [0,T]$, $T = 1000$, and forward-diffusion noise was added to one
randomly selected spatial quarter of the sequence, the same quarter in every image of the sequence
(\textbf{Fig.~\ref{fig:architecture}a}, lower block). The binary quarter mask and the timestep were retained. The forward
process follows the denoising diffusion probabilistic model formulation\cite{ho2020denoising}:

\begin{equation}
    \boldsymbol{x}_t = \sqrt{\bar{\alpha}_t}\, \boldsymbol{x}_0
    + \sqrt{1 - \bar{\alpha}_t}\, \boldsymbol{\epsilon},
    \quad \boldsymbol{\epsilon} \sim \mathcal{N}(\mathbf{0}, \mathbf{I})
    \label{eq:forward_diffusion}
\end{equation}

\noindent where $\bar{\alpha}_t = \prod_{s=1}^{t}(1 - \beta_s)$ and $\beta_s$ follows a linear
variance schedule\cite{ho2020denoising}, here spanning $\beta \in [10^{-5}, 10^{-3}]$. The model is thereby required to restore the corrupted quarter from the three clean quarters and from
$\boldsymbol{p}$, rather than to copy the input. This masked formulation follows the patch-based normative modeling principle established for
anatomical brain MRI\cite{behrendt2024patched}, applied here at quarter-image granularity and
extended from a single anatomical contrast to a six-image non-steady-state molecular ST sequence
conditioned on the excitation parameters.

\noindent\textbf{Embedding.} The noised sequence was divided into 100 non-overlapping spatial
patches; the 6 co-located $4 \times 4$ patches of the sequence were stacked and linearly projected
into a single token, giving 100 tokens of the embedded tissue response (\textbf{Fig.~\ref{fig:architecture}a})\cite{dosovitskiy2020image}. A timestep embedding was obtained by mapping the normalized
timestep $t/T$ through a two-layer MLP with GELU activation to the patch-embedding dimension and
adding it element-wise to the image tokens only, conditioning the model on the noise level. In
parallel, $\boldsymbol{p}$ was mapped through a fully connected layer to an RF excitation embedding. The tissue-response and RF-excitation embeddings
were concatenated into a single token sequence.

\noindent\textbf{Encoder and decoder.} The concatenated tokens were passed through a transformer
encoder\cite{vaswani2017attention} with embedding dimension 256, MLP size 1024, four layers and four
attention heads. A five-layer convolutional decoder then projected the encoder output back to the
input shape: two $3 \times 3$ convolutional layers expand the feature representation, each followed
by batch normalization\cite{ioffe2015batch} and ReLU; an up-sampling step and two further
$3 \times 3$ convolutional layers with batch normalization and ReLU restore the $40 \times 40$
spatial resolution; and a final $1 \times 1$ convolution followed by a sigmoid yields the
reconstructed sequence $\hat{\boldsymbol{x}} \in \mathbb{R}^{6 \times 40 \times 40}$. The model has
approximately 3.9M trainable parameters and requires approximately 17.8 GFLOPs for a single forward
pass at batch size 1.

\noindent\textbf{Training objective.} The network was optimized with a combination of the structural
similarity index measure (SSIM)\cite{wang2004image}, computed over the entire sequence, and an $L_1$ term computed only
within the noised quarter:

\begin{equation}
    \mathcal{L}(x, \hat{x}) = \lambda_1 \left(1 -
    \frac{2\mu_x \mu_{\hat{x}} + C_1}{\mu_x^2 + \mu_{\hat{x}}^2 + C_1} \cdot
    \frac{2\sigma_{x\hat{x}} + C_2}{\sigma_x^2 + \sigma_{\hat{x}}^2 + C_2}
    \right)
    + \lambda_2 \frac{1}{N_m} \sum_{i \in N_m} |\hat{x}_i - x_i|
    \label{eq:loss}
\end{equation}

\noindent where $m$ denotes the noised quarter, $N_m$ the number of pixels within it,
$\lambda_1 = 0.6$, $\lambda_2 = 0.4$, $C_1 = 0.0001$, $C_2 = 0.0009$, and $\mu$, $\sigma^2$ and
$\sigma_{x\hat{x}}$ are the mean, variance and covariance, respectively. Training used a single RTX
5000 GPU with a batch size of 32, a learning rate of $10^{-4}$, and the AdamW
optimizer\cite{loshchilov2017decoupled} with a weight decay of 0.03 and standard $\beta_1$ and
$\beta_2$, and converged in 98 epochs over 12 hours.

\noindent\textbf{Inference.} At inference (\textbf{Fig.~\ref{fig:architecture}b}) the procedure is identical except in two
respects. First, the timestep is fixed at $t_{\mathrm{inf}} = T/4 = 250$, selected on the validation
set. Second, the four quarters are processed in turn: in each of four passes a different quarter is
noised and reconstructed, and only that quarter's reconstruction is retained. The four
denoised quarters are subsequently recombined into a single reconstructed sequence. Every voxel is therefore
reconstructed from context rather than from itself.

\noindent\textbf{Outputs.} Two quantities are derived from the comparison between the input and
reconstructed sequences. The anomaly map is the voxel-wise absolute ($L_1$) difference between them,
computed per image to preserve the spatial structure of the error, and averaged across the $M=6$
images to yield a single mean anomaly map per sample. The anomaly score is the normalized root mean
square error (NRMSE) between the input and reconstruction, computed per image and averaged across
the six images into one value per subject and protocol. Before both computations the brain mask was
morphologically eroded by 3 pixels to  remove boundary effects, and the
NRMSE and anomaly maps were restricted to the eroded mask.

\noindent\textbf{Human adaptation.} For the human data, we retrained the same framework with the following minimal changes. We used only the saturation frequency offset and saturation power as input parameters, yielding a $2 \times 6$ parameter tensor, since we used only the rapid MT protocol (which yielded the best results in the preclinical study). To accommodate the larger $128 \times 128$ image size, we set the batch size to 16, the patch size to 8, the embedding dimension to 768, the MLP dimension to 3072, and the number of transformer layers to 3. Training used nine A5000 Nvidia GPUs.

\noindent\textbf{Computational cost.} Inference required 14.79 $\pm$ 11.91 ms for a single mouse image, 85.37 $\pm$ 6.61 ms for a single human 2D slice image, and 2.28 $\pm$ 1.86 s for a complete 3D human brain analysis, measured on an NVIDIA GeForce RTX 3060 GPU.

\subsection{Performance and statistical analysis.}
AUROC and AUPRC were computed separately for each protocol, and for the four protocols combined by averaging the NRMSE of each animal across protocols into a single per-animal score before computing the metrics, for each anomaly class against the healthy test group and for all anomaly classes combined. The evaluation was repeated across 20 inference seeds, with results reported as the mean $\pm$ standard deviation across seeds. Group differences in NRMSE were tested separately for each protocol, using the anomaly scores of a representative inference seed (closest to the mean of all seeds in terms of AUROC across all categories). Because the ten healthy mouse test scans are the pre-induction baselines of the MS and PD mice, the post-induction scores of these ten mice were compared with their own baselines using an exact two-sided Wilcoxon signed-rank test\cite{wilcoxon1945} (MS and PD pooled; ten matched pairs). The brain cancer and aging mice have no baseline scans and are independent of the healthy group; they were therefore pooled and compared with the ten healthy scans using an exact two-sided Mann--Whitney U test\cite{mann1947test} (10 versus 8; 10 versus 7 for guanidinium). The two p-values are reported separately (\textbf{Fig.~\ref{fig:nrmse}}). Confidence intervals (95\%) for AUROC and AUPRC were obtained by percentile bootstrap (20,000 resamples) with the animal as the resampling unit, so that the baseline and post-induction scans of each induced animal were resampled together. All tests were computed with SciPy\cite{2020SciPy-NMeth}.

The anomaly score of each test participant in the human cohort is defined as the NRMSE between the input and the normative reconstruction, averaged over that participant's slices, and AUROC and AUPRC were computed at the subject level for healthy versus each class and versus all anomalies. Ninety-five percent confidence intervals were obtained by percentile bootstrap over participants (20,000 resamples) using the anomaly scores of the representative inference seed, with healthy participants and patients resampled separately so that every resample contained both classes. At evaluation, the brain mask was eroded by 3 pixels to exclude edge artifacts at the zero-padding boundary, and the NRMSE was computed over the remaining brain voxels.

\subsection{Comparison to alternative models}
The proposed framework was compared with three diffusion-based UAD methods: MAD-AD\cite{beizaee2025mad}, Ano-DDPM\cite{wyatt2022anoddpm}, and ANDi\cite{frotscher2025unsupervised}. All three were developed for, and evaluated on, anatomical human brain MRI in their original publications, and were retrained here using mouse data, for a fair comparison. Two comparative studies were performed. In the first, we retrained the three methods on $T_2$-weighted images from the same mice, using the same training, validation, and test animals as the ST data, and we retrained the proposed framework on the same $T_2$-weighted data with no parameter tensor input. In the second, we retrained the three methods on the same ST data as the proposed framework, using the six-image ST sequence but not the excitation parameters, which their architectures do not accept. They were evaluated on the rapid MT protocol alone and on the four protocols combined.

For all methods, the inference seed determines the realization of the forward-diffusion noise applied at inference (for the proposed framework, $t_{\mathrm{inf}} = 250$, applied to each image quarter in turn), and every method was evaluated with the same 20 inference seeds. For each seed, scan-level anomaly scores were recomputed, and AUROC and AUPRC were derived per protocol and per anomaly class. Results are reported as the mean $\pm$ standard deviation across seeds, together with the mean $\pm$ standard deviation of the per-seed AUROC difference between the proposed framework and each alternative, and the number of seeds in which the proposed framework ranked first. Conditions separated perfectly in every seed (zero variance) are reported as such and excluded from the variability statistics.

\end{methods}

\section*{Acknowledgments}
The authors thank Tony Stöcker and Rüdiger Stirnberg for their help with the 3D EPI readout. This project was funded by the European Union (ERC, BabyMagnet, project no. 101115639). However, the views and opinions expressed are those of the authors only and do not necessarily reflect those of the European Union or the European Research Council. Neither the European Union nor the granting authority can be held responsible for them. This work was supported by the Israel Science Foundation (grant No. 1030/25), the German Research Foundation (DFG, grant No. 559955894), and the Michael J. Fox Foundation for Parkinson’s Research (grant ID MJFF-028098). The study was supported in part by Moonshot-Med, a joint grant program from Tel Aviv University and Clalit Innovation, the innovation arm of Clalit Health Services, and the Edmond J. Safra Center for Bioinformatics at Tel Aviv University. 

\section*{Author contributions}
Conceptualization: S.I. and O.P., deep learning methodology: S.I. and O.P., animal model establishment: A.E., M.R., R.B.C., H.S., O.Z., D.F.M., preclinical MRI acquisition: A.E., M.R., R.B.C., H.S., and O.P., clinical MRI protocol engineering: S.W., M.Z., and O.P., clinical MRI data acquisition and/or subject recruitment: M.R., M.Z., S.G.Z., D.B.B., S.Y.K, R.S.H, and O.P., ablation studies, comparison with state of the art, and data analysis: S.I., manuscript writing: S.I. and O.P., manuscript reviewing and editing: all authors, supervision: O.P.

\section*{Competing interests}
O.P. and S.I. applied for a patent related to the proposed framework (PCT/IL2025/050688). O.P. is an Editorial Board Member for \textit{Communications Engineering} but was not involved in the editorial review or peer review, nor in the decision to accept/reject this article.  All other authors declare no competing interests.

\subsection {Disclosure of artificial intelligence (AI) assisted technologies.}
Claude Code (Anthropic) was used for occasional code-writing and text structure, organization, and refinement. The authors drafted, reviewed, edited, and validated all content and take full responsibility for the final code and text.

\subsection{Data availability}
Animal image data used to create all manuscript figures, as well as sample human data, will be made available upon acceptance in a dedicated GitHub repository and in Zenodo.

\subsection{Code Availability}
The clinical MRI acquisition protocols used in this work are available in .seq format in the pulseq CEST open library: https://github.com/kherz/pulseq-cest-library/tree/master/seq-library\cite{herz2021pulseq}. Preclinical acquisition protocols can be reproduced using the definitions provided in https://github.com/momentum-laboratory/molecular-mrf\cite{vladimirov2025quantitative}. Trained anomaly detection networks and relevant code will become available upon publication in a dedicated GitHub repository and in Zenodo.

\noindent \textbf{References}
\bibliography{Paper_Bib}

\end{document}